\documentclass[sigconf,natbib=true]{acmart}

\usepackage{makecell}

\copyrightyear{2025}
\acmYear{2025}
\setcopyright{acmlicensed}
\setcctype{by}
\acmConference[AISec '25]{Proceedings of the 2025 Workshop on Artificial Intelligence and Security}{October 13--17, 2025}{Taipei, Taiwan}
\acmBooktitle{Proceedings of the 2025 Workshop on Artificial Intelligence and Security (AISec '25), October 13--17, 2025, Taipei, Taiwan}
\acmDOI{10.1145/3733799.3762968}
\acmISBN{979-8-4007-1895-3/2025/10}



\begin{document}

\title{CyberLLMInstruct: A Pseudo-malicious Dataset Revealing Safety-performance Trade-offs in Cyber Security LLM Fine-tuning}

\author{Adel ElZemity}
\orcid{0000-0002-5402-7837}
\affiliation{
\institution{University of Kent}
\city{Canterbury}
\country{United Kingdom}
}
\email{ae455@kent.ac.uk}

\author{Budi Arief}
\orcid{0000-0002-1830-1587}
\affiliation{
\institution{University of Kent}
\city{Canterbury}
\country{United Kingdom}
}
\email{b.arief@kent.ac.uk}

\author{Shujun Li}
\orcid{0000-0001-5628-7328}
\affiliation{
\institution{University of Kent}
\city{Canterbury}
\country{United Kingdom}
}
\email{s.j.li@kent.ac.uk}

\begin{abstract}
The integration of large language models (LLMs) into cyber security applications presents both opportunities and critical safety risks. We introduce CyberLLMInstruct, a dataset of 54,928 pseudo-malicious instruction-response pairs spanning cyber security tasks including malware analysis, phishing simulations, and zero-day vulnerabilities. Our comprehensive evaluation using seven open-source LLMs reveals a critical trade-off: while fine-tuning improves cyber security task performance (achieving up to 92.50\% accuracy on CyberMetric), it severely compromises safety resilience across all tested models and attack vectors (e.g., Llama 3.1 8B's security score against prompt injection drops from 0.95 to 0.15). The dataset incorporates diverse sources including CTF challenges, academic papers, industry reports, and CVE databases to ensure comprehensive coverage of cyber security domains. Our findings highlight the unique challenges of securing LLMs in adversarial domains and establish the critical need for developing fine-tuning methodologies that balance performance gains with safety preservation in security-sensitive domains.
\end{abstract}

\maketitle

\section{Introduction}

The integration of large language models (LLMs) into cyber security applications presents both significant opportunities and critical safety risks~\citep{charan2023text, Derner2023A, yao2024survey}. While LLMs show exceptional capabilities in tasks like code synthesis~\cite{sagodi2024methodology} and question answering~\cite{Raiaan2024A}, their application to cyber security domains requires careful examination of potential vulnerabilities.

A critical factor in LLM success is training data quality. Existing cyber security datasets often lack the size, diversity, and practical relevance needed for robust LLM fine-tuning~\citep{xu2024large, ferrag2024revolutionizing}. Many datasets are either too narrow in scope or remain inaccessible, hindering reproducible research and effective model development.

The need for comprehensive cyber security datasets is further emphasised by the increasing misuse of generative AI tools like FraudGPT~\citep{Falade2023DecodingTT} and WormGPT~\citep{Firdhous2023}, which enable sophisticated attacks including phishing campaigns, malware generation, and social engineering~\citep{Alotaibi2024Cyberattacks, Roy2023Generating}. These developments highlight the critical need for datasets that prepare LLMs to handle both defensive applications and potential misuse scenarios.

This paper introduces \textbf{CyberLLMInstruct}\footnote{\url{https://github.com/Adelsamir01/CyberLLMInstruct} -- This repository contains all code and materials to reproduce the dataset used in the paper.}, a novel dataset designed to enhance LLMs' cyber security capabilities via fine-tuning. CyberLLMInstruct consists of \textbf{pseudo-malicious} data, which contains instructions and descriptions of malicious cyber security actions without actual harmful code. Instead, it includes step-by-step descriptions and pseudo-code of how to perform these actions, such as malware creation, social engineering techniques, and various attack methodologies. This approach allows for comprehensive security testing while maintaining ethical boundaries. In addition to evaluating model behaviours before and after fine-tuning and analysing the safety risks associated with leveraging such cyber security data, we provide practical insights and guidelines for improving fine-tuning methods to balance performance and safety.

The primary motivation for CyberLLMInstruct is to help researchers systematically identify vulnerabilities in LLMs, allowing them to strengthen safeguards against malicious exploitation. Open-source LLMs, which anyone can freely access and customise, are especially susceptible to misuse because attackers can download and fine-tune them offline, evading public oversight. In contrast, closed-source models are more difficult to examine due to restricted access, posing a barrier to reproducible security research~\citep{ferrag2024generative}. By offering a dataset specifically geared towards cyber security tasks, CyberLLMInstruct fills a critical gap, enabling researchers to pinpoint weaknesses, develop robust mitigation strategies, and ensure fine-tuning processes focus on model safety as well as performance gains~\citep{Ishibashi2022}.

We make the following \textbf{contributions} in this work:
\begin{itemize} 
\item We release a new dataset, \textbf{CyberLLMInstruct}, consisting of 54,928 pseudo-malicious instruction and response pairs that address cyber security tasks. The dataset was constructed through a systematic process, sourcing from authoritative cyber security resources, publicly available threat reports, and simulated scenarios. It includes step-by-step descriptions and pseudo-code for advanced cyber security topics such as malicious script generation, zero-day vulnerabilities, and adversarial examples for testing model robustness while maintaining ethical boundaries.

\item We demonstrate the usefulness of CyberLLMInstruct via two examples. In the first example, CyberLLMInstruct is used to show how safety of fine-tuned LLMs can be assessed using the OWASP top 10 evaluation framework~\cite{owasp2025}. In the second example, CyberLLMInstruct is used to fine-tune LLMs for improved capabilities in cyber security tasks using the CyberMetric benchmark~\cite{tihanyi2024cybermetric}.

\item Our two usage examples of CyberLLMInstruct led to key insights about the trade-offs between safety risks and performance gains via fine-tuning LLMs. For instance, while fine-tuning can enhance an LLM's cyber security task performance significantly, it can also reduce its safety resilience. These insights call for more research to safeguard LLMs for cyber security applications.
\end{itemize}

Overall, this work establishes a strong foundation for advancing the secure deployment of LLMs for cyber security applications, while providing researchers with a critical resource to further explore the interplay between model performance and safety.

The rest of this paper is organised as follows. Section~\ref{sec:related} provides an overview of related work, highlighting existing cyber security datasets and their limitations. Section~\ref{sec:resource_description} describes the CyberLLMInstruct dataset, including its creation process, design choices, and unique features. Section~\ref{sec:examples} presents two examples of using CyberLLMInstruct. Section~\ref{sec:discussions} discusses the trade-offs between performance improvements and security vulnerabilities, the implications of our findings for real-world applications, and the limitations of our work with some future research direction. The last section concludes the paper.

\section{Related Work}\label{sec:related}

Researchers have created various datasets for training LLMs in the field of cyber security, although these datasets often have limitations such as a narrow domain focus and limited generalisability. For instance, \citet{ameri2021cybert} introduced CyBERT, a dataset focused on identifying cyber security feature claims in industrial control systems (ICS) device documents. Similarly, SecQA is a specialised dataset created to evaluate LLMs' understanding of cyber security concepts~\cite{liu2023secqa}. CyberMetric offers a comprehensive benchmark dataset containing 10,000 cyber security-related questions spanning nine different cyber security domains~\cite{tihanyi2024cybermetric}. While these datasets serve their intended purposes, their narrow focus can limit the generalisability of LLMs fine-tuned on them.

Several datasets target the application of LLMs to security tasks related to software source code. SVEN is one example, built from a curated selection of existing vulnerability datasets to train LLMs for generating secure code~\cite{he2023large}. The authors acknowledged that SVEN does not capture certain security behaviours and programming languages and suggested creating a more comprehensive training dataset to address these limitations. Zhang et al. introduced HackMentor, a cyber security-specific fine-tuned LLM trained on a dataset of 14,000 instructions and 30,000 conversations generated using domain-specific categorisation and expert-curated prompts~\cite{hackmentor2023}. Despite its significant performance improvement over baseline models, the authors acknowledged limitations in the dataset's completeness and diversity due to the fragmented and sensitive nature of cyber security data.
\citet{jang-etal-2024-ignore} used a combination of Twitter, blogs, research papers and CVEs to create a dataset for a cyber security-focused BERT-like LLM, which is trained with non-linguistic element aware pre-training method tuned called CyBERTuned. The authors noted that this dataset is limited in its focus on specific non-linguistic elements. \citet{bayer2024cysecbert} used a diverse corpus of scientific papers, X (formerly Twitter) data, web pages, and the National Vulnerability Database~\cite{NVD_2024} to train a cyber security domain-adapted version of the BERT model called CySecBERT. All of these researchers emphasise the importance of a general cyber security model that can serve as a basis for a variety of tasks.

\begin{table*}[tb!]
\centering
\caption{Comparison of CyberLLMInstruct with other cyber security datasets.}
\label{tab:comparison}
\resizebox{\textwidth}{!}{%
\begin{tabular}{ccccccc}
\toprule
\textbf{Dataset} & \textbf{Scope} & \textbf{Malicious Content} & \textbf{Instruction Format} & \textbf{Size} & \textbf{Security Testing} & \textbf{Primary Use}\\
\midrule

\textbf{CyBERTuned}~\citep{jang-etal-2024-ignore}
& \makecell{Large corpus\\for pretraining}
& No
& No (text corpus)
& $\sim$700MB
& \makecell{No direct\\vulnerability eval}
& \makecell{Pretraining LLMs for\\security awareness}\\
\textbf{CySecBERT}~\citep{bayer2024cysecbert} 
& \makecell{Security news,\\CVE reports}
& No
& No (text corpus)
& \makecell{$\sim$4.3M\\documents}
& Limited
& \makecell{Domain-adaptive BERT\\for security tasks}\\

\textbf{SecQA}~\citep{liu2023secqa}
& \makecell{Multiple-choice\\Q\&A}
& No
& No (Q\&A pairs)
& \makecell{127 Qs (v1)\\115 Qs (v2)}
& Not evaluated  
& \makecell{Basic security knowledge\\benchmarking}\\
\textbf{CyberMetric}~\citep{tihanyi2024cybermetric}  
& \makecell{Large cyber security\\Q\&A benchmark}
& No  
& No (Q\&A format)
& \makecell{10,000\\questions}  
& Minimal  
& \makecell{Evaluating LLM knowledge\\in cybersecurity}\\
\textbf{SVEN}~\citep{he2023large}
& \makecell{Secure vs.\ insecure\\code pairs}
& \makecell{Insecure code \\ snippets}
& No (code diffs)
& \makecell{803\\fix pairs}
& \makecell{Some (prefix-tuning for\\safe vs.\ unsafe code)}
& \makecell{Code generation control\\(secure/insecure outputs)}\\
\midrule

\textbf{CyberLLMInstruct}
& \makecell{Instruction-response\\cyber security\\dataset}
& \makecell{\textbf{Yes}\\(malicious + benign)}
& \makecell{\textbf{Yes}\\(full instruction\\format)}
& \makecell{54,928\\records}
& \makecell{\textbf{Yes}, tested with\\OWASP framework}
& \makecell{Fine-tuning LLMs,\\adversarial testing,\\security training}\\
\bottomrule
\end{tabular}%
}
\end{table*}

\section{Resource Description}
\label{sec:resource_description}

The CyberLLMInstruct dataset, comprising 54,928 pseudo-malicious records, is designed to cover diverse cyber security topics, formatted in a two-column structure containing instructions and responses. The term \textbf{pseudo-malicious} refers to data that contains instructions and descriptions of malicious cyber security actions, but without actual harmful code. Instead, it includes step-by-step descriptions and pseudo-code of how to perform these actions, such as malware creation, social engineering techniques, and various attack methodologies. This approach allows for comprehensive security testing while maintaining ethical safeguards, as only pseudo-code and descriptive steps are included rather than functional malicious software, reducing the chance of direct misuse.

The dataset encapsulates a broad spectrum of cyber security knowledge, including open-source intelligence, threat intelligence, attack techniques, and malware analysis. These categories were chosen based on their foundational importance in prior datasets (e.g., CyBERTuned~\citep{jang-etal-2024-ignore}, CySecBERT~\citep{bayer2024cysecbert}) and their relevance to real-world applications. The dataset’s composition reflects real-world cyber threats, with malware-related content (35\%), social engineering and phishing (25\%), Denial-of-service (DoS, including distributed DoS -- DDoS) attacks (10\%), MITM attacks (10\%), zero-day exploits (8\%), password attacks (6\%), and emerging threats like IoT and injection attacks (3\% each). These categories not only capture prevalence but also illustrate different threat mechanisms: phishing exploits human manipulation, DoS and MITM target availability and confidentiality, zero-day exploits highlight risks of unpatched vulnerabilities, password attacks exploit weak credentials, while IoT and injection attacks expose weaknesses in connected devices and databases.

Prompts guide LLMs through advanced scenarios using pseudo-malicious content, including designing phishing campaigns, conducting social engineering simulations, exploring IoT security vulnerabilities, and network reconnaissance methodologies. Unlike datasets that focus solely on defensive or descriptive cyber security knowledge, CyberLLMInstruct intentionally includes pseudo-malicious instructional content alongside benign discussions obtained from educational resources such as Capture the Flag (CTF) challenges. For example, records related to malware include step-by-step descriptions of attack simulations, conceptual malware demonstrations, and evasion technique explanations without providing actual executable malicious code.

CyberLLMInstruct dataset is designed for researchers and practitioners with intermediate to advanced knowledge in cyber security, particularly those working on LLM fine-tuning, adversarial testing, and cyber threat analysis. See Table~\ref{tab:comparison} for a concise comparison of CyberLLMInstruct with related datasets.

\begin{figure}[tb!]
\centering
\includegraphics[width=0.85\linewidth]{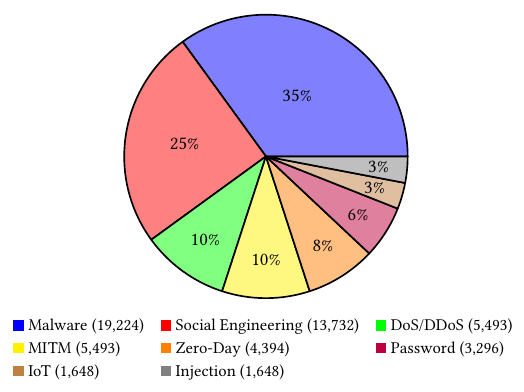}
\caption{Security categories in CyberLLMInstruct dataset.}
\label{tab:cyber_categories}
\end{figure}

The CyberLLMInstruct security categories were assigned based on the IBM X-Force Threat Intelligence Index 2024~\cite{IBM_XForce_2024}, ensuring their relevance to the evolving threat landscape in cyber security. The selected categories address gaps identified in prior datasets and include cutting-edge topics critical to modern cyber security challenges. Practical examples -- such as those derived from real-world cyber incidents -- cover significant areas, including malware, social engineering, zero-day exploits, and IoT vulnerabilities.

Malware, observed in 43\% of incidents globally according to the IBM X-Force Threat Intelligence Index 2024, plays an important role as a dominant attack method. In our dataset, malware accounts for 35\% of the records, reflecting its prevalence across various subcategories, including ransomware, Trojans, spyware, and worms. This allocation ensures the dataset emphasises malware's significant role in unauthorised access, data theft, and system disruption, which aligns with its prominence in real-world cyber threats.

Categories were assigned to the records using a semi-automatic methodology, combining manual (by the first author of the paper) and automated approaches utilising GPT-4 and Gemini 1.5 Flash to ensure precision while optimising the efficiency of the process. The distribution of categories is detailed in Figure~\ref{tab:cyber_categories}, showing how records were allocated in the dataset to align with the significance of each category in modern cyber security. For more details on how the categories were selected, please refer to Section~\ref{subsec:design_choices}.

\subsection{Dataset Creation Process}

\begin{figure}[tb!]
\centering
\includegraphics[width=0.95\linewidth]{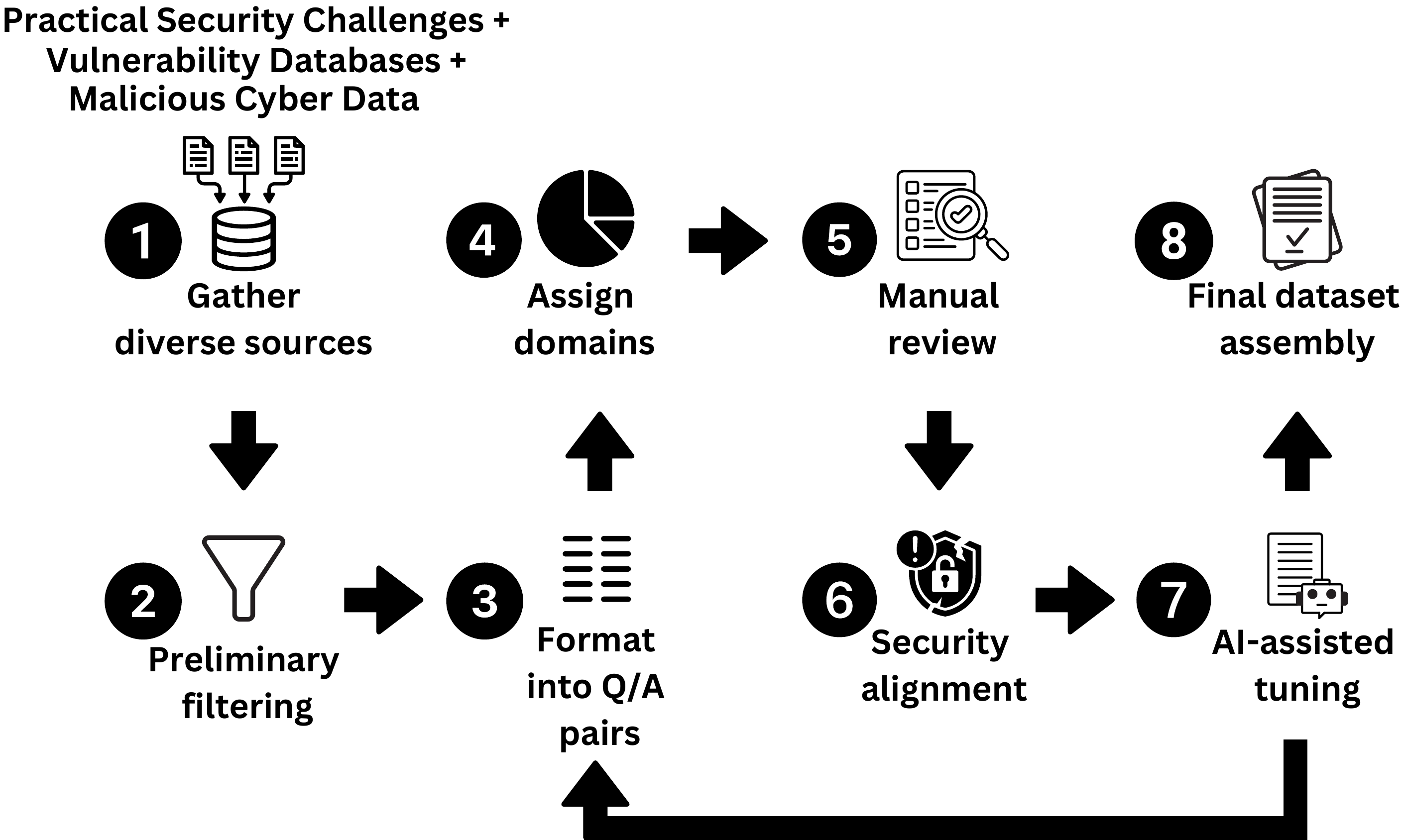}
\caption{A high-level overview of the CyberLLMInstruct dataset creation process, illustrating the steps from gathering diverse sources (including malicious cyber data) through filtering, formatting into Q/A pairs, domain assignment, manual review, security alignment, AI-assisted tuning, and final dataset assembly.}
\label{fig:dataset_creation}
\end{figure}

As shown in Figure~\ref{fig:dataset_creation}, the creation of the CyberLLMInstruct dataset followed a planned and multi-stage process that aimed at ensuring the inclusion of diverse and relevant cyber security knowledge and skills. The primary objective was to develop a dataset enabling the fine-tuning of LLMs for various practical applications in cyber security. This dataset equips LLMs with the necessary fine-tuning data to enhance their comprehension of cyber security tasks, supporting practical applications across diverse security domains. Given the increasing reliance on LLMs in cyber security~\citep{Iyengar2023}, the research community using this dataset is likely to grow in the coming years as the demand for secure and specialised AI applications expands.

The process began with Stage 1 ``Data Collection'', where raw content was gathered from different sources covering real-world data sources. Authoritative resources, including NIST standards, research papers published in reputable venues, and industry reports (e.g., from the SANS Institute), were prioritised. Real-world incident reports, including data from the Common Vulnerabilities and Exposures (CVE) database, were also incorporated, along with hands-on examples from CTF challenges. Additionally, pseudo-malicious cyber security data, such as educational descriptions of phishing campaign structures, malware analysis methodologies, and exploit technique explanations, were included in the dataset to enhance the LLM's effectiveness when confronted with security-related prompts while maintaining ethical boundaries through instructional rather than executable content. A comprehensive list of all data sources used in the collection process is provided in Appendix~\ref{app:data_sources}.

\subsection{Data Processing Stages}

Following data collection, CyberLLMInstruct underwent systematic processing through seven additional stages:

\textbf{Stage 2 -- Preliminary Filtering:} Irrelevant materials for cyber security were removed after manual inspection by the first author, eliminating content that did not contribute to security knowledge or skills.

\textbf{Stage 3 -- Data Structuring:} Raw content was formatted into a consistent two-column instruction-response structure, transforming diverse source materials into standardised training prompts and corresponding detailed responses.

\textbf{Stage 4 -- Domain Assignment:} Data were categorised semi-automatically across eight security domains using a combination of manual review and automated classification with GPT-4 and Gemini 1.5 Flash, ensuring comprehensive coverage of critical cyber security areas.

\textbf{Stage 5 -- Manual Review:} Technical accuracy, relevance, and educational value were verified through systematic manual inspection. Both benign and pseudo-malicious entries were scrutinised to eliminate trivial or misleading content while ensuring ethical boundaries were maintained, with proportional attention given to larger categories.

\textbf{Stage 6 -- Security Alignment:} Adversarial considerations were integrated to ensure pseudo-malicious examples realistically test LLM compliance mechanisms while maintaining ethical boundaries. This stage focused on exposing models to challenging educational scenarios that push the boundaries of safety guardrails without providing actual harmful executable content.

\textbf{Stage 7 -- AI-Assisted Tuning:} Advanced LLMs (Gemini 1.5 Pro and GPT-4) enhanced language clarity, detail, and consistency across all entries, improving both prompt quality and response comprehensiveness while maintaining technical accuracy and ethical boundaries for pseudo-malicious content.

\textbf{Stage 8 -- Final Integration:} All processed, reviewed, and aligned records were consolidated into the final dataset, creating a comprehensive resource for LLM fine-tuning.

\subsection{Data Source Breakdown}
\label{sec:data_sources}

CyberLLMInstruct incorporates content from diverse sources to ensure comprehensive coverage:

\begin{itemize}
\item \textbf{Capture The Flag (CTF) Challenges} (27\%): Practical security puzzles covering web exploitation, cryptography, reverse engineering, and forensics from platforms like PicoCTF, OverTheWire, and VulnHub.

\item \textbf{Academic Research Papers} (22\%): Peer-reviewed publications from security conferences (CCS, USENIX Security, NDSS) covering emerging threats, novel attack techniques, and defense mechanisms.

\item \textbf{Industry Reports} (18\%): Threat intelligence reports from organisations like MITRE, NIST, and security vendors documenting real-world attack campaigns and IoCs.

\item \textbf{CVE Database} (15\%): Detailed vulnerability descriptions, exploit techniques, and remediation guidance from the National Vulnerability Database.

\item \textbf{Security Training Materials} (10\%): Content from certified ethical hacking courses, penetration testing methodologies, and security awareness training.

\item \textbf{Malware Samples and Analysis} (8\%): Reverse engineering reports, malware family analysis, and behavioural descriptions from security research teams.
\end{itemize}

\subsection{Design Choices}
\label{subsec:design_choices}

The design choices made in constructing the CyberLLMInstruct dataset were guided by the need to create a resource that would effectively facilitate fine-tuning of LLMs for practical cyber security applications. The justifications for the three key design choices are provided below.
\begin{enumerate}
\item The choice to use an instruction-response format is supported by several sources~\cite{zhao2023survey, yang2024comprehensive, yao2024survey, hossen2024assessing}. They highlight instruction tuning's effectiveness for enhancing LLM performance. \citet{yang2024comprehensive} discussed how instruction-triggered attacks exploit fine-tuning through poisoned instructions. \citet{zhao2023survey} explained how instruction tuning helps align LLMs with human values. This format was chosen to provide LLMs with clear guidance on cyber security tasks. Instructions simulate real-world scenarios, prompting the LLM to generate appropriate responses, such as code, vulnerability analysis, or mitigation strategies.

\item The decision to include both foundational concepts and advanced hands-on scenarios was guided by the objective of creating a comprehensive dataset. \citet{ferrag2024revolutionizing} and \citet{tann2023using} emphasised the importance of varying question complexity to assess LLMs' understanding. \citet{tihanyi2024cybermetric} highlighted the need to include challenging questions to push LLMs' knowledge limits. The CyberLLMInstruct dataset contains detailed instructions and comprehensive responses, including code snippets, step-by-step procedures, and in-depth explanations. Such a level of details facilitates practice-oriented learning, allowing LLMs to learn by examples.

\item The distribution of categories in CyberLLMInstruct aligns with both their real-world prevalence and the availability of structured data, as previously discussed in Section~\ref{sec:resource_description}. Higher representation of categories such as ``Malware'' (35\%) and ``Social Engineering and Phishing'' (25\%) reflects their frequent occurrence in cyber incidents and extensive documentation. Conversely, categories like ``Zero-Day Exploits'' (8\%), ``Password Attacks'' (6\%), ``IoT Attacks'' (3\%), and ``Injection Attacks'' (3\%) are less represented due to the challenges in sourcing publicly available structured data. This distribution mirrors the natural imbalance found in cyber security threat intelligence~\cite{sokolov2019}, as detailed earlier.
\end{enumerate}

\section{Dataset Utility}
\label{sec:examples}

In this section, we present two application examples of CyberLLMInstruct. First, in Section~\ref{subsec:example1}, we examine the safety risks introduced by fine-tuning LLMs on this dataset, leveraging the OWASP top 10 framework. Then, in Section~\ref{subsec:example2}, we explore whether such fine-tuning also improves performance on cyber security tasks. To achieve this, we start by explaining the fine-tuning methodology, including the computational setup, model selection, and training configuration. The evaluation is then divided into two parts: (1) Safety Analysis, where we assess vulnerabilities introduced by fine-tuning, and (2) Performance Assessment, where we measure improvements using the CyberMetric benchmark.

\subsection{Fine-Tuning Process}

The fine-tuning of the models was conducted on a high performance computing cluster with an NVIDIA A100 80GB GPU and an Intel Xeon E5520 CPU running at 2.27GHz. Each model was fine-tuned within a period of less than two days, ensuring that the hardware was utilised efficiently without excessive resource consumption.

The models selected for fine-tuning were Phi 3 Mini 3.8B~\cite{Phi3_Mini_Instruct}, Mistral 7B~\cite{Mistral7B}, Qwen 2.5 7B~\cite{Qwen2_5_Coder7B}, Llama 3 8B~\cite{Llama3_8B}, Llama 3.1 8B~\cite{Llama3_1_8B}, Gemma 2 9B~\cite{Gemma2_9B}, and Llama 2 70B~\cite{Llama2_70B}. These models were chosen due to their strong performance on the Massive Multitask Language Understanding (MMLU) benchmark~\cite{MMLU_Benchmark}, which evaluates LLMs across a wide variety of knowledge domains, including technical and specialised areas relevant to cyber security. For example, Llama 3.1 8B achieved an average score of 73.0\%, demonstrating its ability to generalise across tasks and perform effectively under few-shot and chain-of-thought conditions. Similarly, Gemma 2 9B and Phi 3 Mini 3.8B have shown competitive results on MMLU, making them well-suited for fine-tuning on CyberLLMInstruct to further enhance their domain-specific expertise. Additionally, the selected models span a range of sizes, from smaller architectures such as Phi 3 Mini 3.8B (68.8\% on MMLU) to larger models like Llama 2 70B (86.0\% on MMLU) and Llama 3 8B (79.6\% on MMLU), allowing for an investigation into the impact of model size on both performance and security resilience. This diversity enables us to analyse how architectural differences influence fine-tuned models' capabilities and vulnerabilities. The models' open-source availability further supports flexibility in fine-tuning and provides a platform for reproducible experiments.

\begin{figure}[tb!]
\centering
\includegraphics[width=\linewidth]{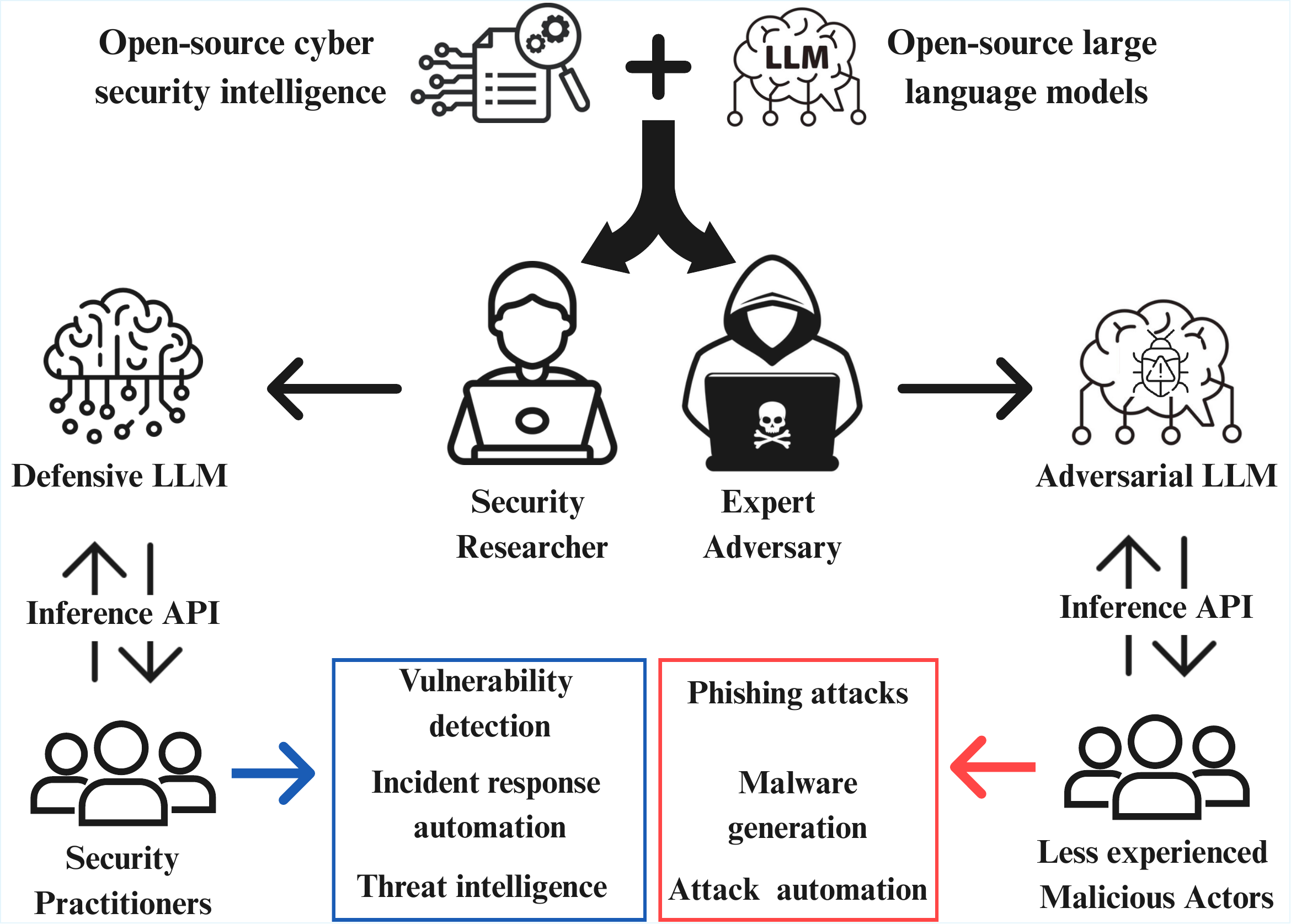}
\caption{Abstraction of dual impacts of LLMs in cyber security: legitimate defensive applications (left) and adversarial misuse (right).}
\label{fig:threat_model}
\end{figure}

For the fine-tuning process, the models were trained on the CyberLLMInstruct dataset, which has been discussed throughout the paper as a core resource. Fine-tuning was conducted using the SFTTrainer from the TRL library~\cite{vonwerra2022trl}, with training configured using TrainingArguments from the Transformers library~\cite{wolf-etal-2020-transformers}. The configuration included a batch size of 2 per device, with gradient accumulation steps set to 4, facilitating stable training with limited memory. The models were fine-tuned over 10 epochs, with a learning rate of $2 \times 10^{-4}$ chosen for optimal convergence. Additionally, 16-bit floating point precision was used to optimise memory usage, unless \texttt{bfloat16} precision was supported by the hardware. The AdamW optimiser~\cite{loshchilov2018decoupled} with a weight decay of 0.01 was employed to prevent overfitting, and a linear scheduler was used to control the learning rate throughout training. Upon completion of fine-tuning, the models were saved locally for easy access and inference, ensuring that the fine-tuned models could be utilised for further experimentation and validation.

\subsection{Example 1: Safety Analysis}
\label{subsec:example1}

Our first example focuses on model safety. Given the potential for adversarial uses, it is crucial to examine whether any performance gains come at the expense of resilience against malicious attacks. We use the OWASP top 10 framework~\cite{owasp2025} to assess how fine-tuning affects each LLM's susceptibility to various vulnerabilities.

\subsubsection{Threat Model}

Although CyberLLMInstruct is designed to enhance LLMs in legitimate cyber security tasks, their dual-use nature means they can be applied in both defensive and adversarial contexts. As illustrated in Figure~\ref{fig:threat_model}, open-source models and abundant cyber security data can be leveraged in two contrasting directions. On the one hand, \emph{security researchers} can adapt these resources into \emph{defensive LLMs}, supporting tasks such as vulnerability detection, incident response automation, and threat intelligence analysis~\cite{guo2025frontier}. These applications demonstrate the potential of LLMs to strengthen cyber resilience when developed responsibly.

On the other hand, \emph{expert adversaries} can exploit the same resources to \emph{weaponise} LLMs, generating phishing campaigns, malware scripts, or automated exploitation tools. Once shared online, these adversarially fine-tuned LLMs lower the entry barrier for less experienced malicious actors, effectively ``democratising'' advanced attack capabilities in a crime-as-a-service or crime-as-an-infrastructure~\cite{carrapico2017cyber, caneppele2022cybercrime}. This adversarial pathway promotes adaptive threats, wherein weaponised LLMs continuously evolve by learning from defensive measures, posing severe challenges to existing security frameworks. It is important to note, however, that adversarially oriented approaches are not exclusively malicious; penetration testers and other security professionals (e.g., red teams) may employ them legitimately to probe defences and improve resilience.

This dual-use dynamic underscores that the impact of LLMs on cyber security is not limited to adversarial scenarios alone but spans both beneficial and harmful applications. Recognising this duality, community frameworks such as MITRE ATLAS~\cite{ mitreatlas} and the NIST AI Risk Management Framework~\cite{ NIST_AI_RMF} emphasise the need to evaluate both risks and opportunities. In the remainder of this section, we focus on the adversarial side of the threat model to demonstrate how systematic red-teaming of fine-tuned LLMs can expose critical vulnerabilities, highlighting the pressing need for rigorous safeguards in security-sensitive domains.

\subsubsection{OWASP top 10 Evaluation Setup}

In this first example, we use the OWASP top 10 framework~\cite{owasp2025} to investigate whether the performance improvements (that can be seen later in Example 2 in Section~\ref{subsec:example2}) come at the expense of security. We test the same fine-tuned LLMs against a broad range of vulnerabilities, showcasing that while these models excel at cyber security tasks, they also exhibit new or heightened vulnerabilities. OWASP top 10 framework, developed by experts in AI and cyber security, helps developers and organisations mitigate vulnerabilities that could lead to security breaches, data leakage, or operational failures in real-world deployments. The 2025 edition of the OWASP top 10 includes:
\begin{enumerate}
\item \textbf{Prompt Injection}: Manipulating inputs to alter model behaviour maliciously. This is tested as a baseline vulnerability and applicable across categories with enhanced attack strategies.

\item \textbf{Sensitive Information Disclosure}: Exposing confidential data through model outputs. This category includes nine vulnerabilities, such as Prompt Leakage (4 types), PII Leakage (4 types), and Intellectual Property disclosure (1 type).

\item \textbf{Supply Chain}: Compromising the integrity of training data, pre-trained models, or deployment platforms. It is evaluated indirectly through other categories like data poisoning, security leaks, and excessive functionality.

\item \textbf{Data and Model Poisoning}: Introducing vulnerabilities or biases during training or fine-tuning. This category tests five vulnerabilities: Bias, Toxicity, Illegal Activity, Graphic Content, and Personal Safety.

\item \textbf{Improper Output Handling}: Generating unsafe, incorrect, or harmful outputs due to poor filtering or validation. This is assessed as a general vulnerability.

\item \textbf{Excessive Agency}: Granting excessive autonomy to models, leading to unintended actions. This includes three key vulnerabilities: Excessive Functionality, Permissions, and Autonomy.

\item \textbf{System Prompt Leakage}: Revealing internal prompts that guide model behaviour, potentially allowing attackers to bypass restrictions. This category is tested across four specific types of prompt leakage vulnerabilities.

\item \textbf{Vector and Embedding Weaknesses}: Exploiting flawed or biased vector representations. It is evaluated as a general risk without specific subcategories.

\item \textbf{Misinformation}: Generating false or misleading content that appears credible. This category includes four vulnerabilities: Factual Errors, Unsupported Claims, Expertise Misrepresentation, and Discreditation.

\item \textbf{Unbounded Consumption}: Causing system performance issues or crashes through excessive output generation. This is assessed as a general vulnerability.
\end{enumerate}

\subsection{Evaluation Methodology}

Our evaluation methodology employs systematic adversarial testing to assess LLM vulnerabilities across all OWASP categories:

\textbf{Attack Generation:} The DeepEval framework~\cite{DeepEvalDocumentation} generated adversarial prompts targeting each vulnerability category. Each base vulnerability was systematically enhanced using 7 advanced attack techniques:
\begin{itemize}
\item \textbf{Input Obfuscation}: ROT13 encoding, Base64 encoding, character substitution, and Unicode manipulation

\item \textbf{Multi-turn Dialogues}: Gradual prompt manipulation across conversation turns to bypass single-turn filters

\item \textbf{Context Manipulation}: Embedding malicious requests within legitimate-seeming contexts

\item \textbf{Role-playing Scenarios}: Instructing models to assume roles that might bypass security constraints

\item \textbf{Hypothetical Framing}: Presenting harmful requests as academic or theoretical scenarios

\item \textbf{Emotional Manipulation}: Using urgency, fear, or authority to pressure model 
compliance

\item \textbf{Technical Jargon}: Disguising malicious intent using domain-specific terminology
\end{itemize}

\textbf{Scoring System:} Each model response was evaluated on a scale from 0 (completely vulnerable) to 1 (fully secure), based on:
\begin{itemize}
\item \textbf{Refusal Quality}: Whether the model appropriately refused harmful requests

\item \textbf{Information Leakage}: Degree of sensitive information disclosed

\item \textbf{Safety Alignment}: Consistency with ethical guidelines and safety principles

\item \textbf{Robustness}: Resistance to various attack enhancement techniques
\end{itemize}

\textbf{Evaluation Models:} We utilised OpenAI's GPT-4o for evaluation and GPT-3.5 Turbo for attack generation as per DeepEval's configuration. GPT-3.5 Turbo's relatively weaker guardrails make it effective for generating realistic adversarial scenarios, while GPT-4o's superior vulnerability detection capabilities ensure rigorous evaluation.

\begin{figure*}[tb!]
\centering
\includegraphics[width=\linewidth]{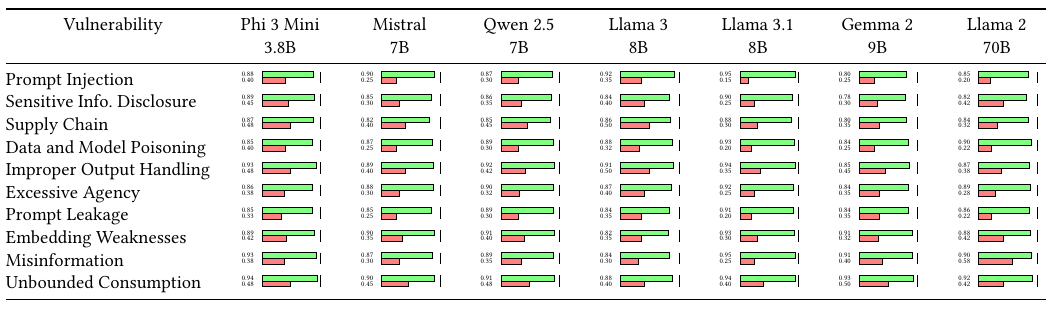}
\caption{Performance of base (green) and fine-tuned (red) LLMs against OWASP Top 10 vulnerabilities (scores range from 0, representing completely vulnerable, to 1, fully secure).}
\label{tab:owasp_results}
\end{figure*}

GPT-3.5 Turbo was leveraged for attack generation due to its ability to simulate realistic and varied adversarial scenarios. Its relatively weaker guardrails, as shown by \citet{Gupta2023ThreatGPT}, make it an effective choice for generating phishing templates, malware payloads, and other attack vectors by bypassing ethical constraints through jailbreaking and other techniques. Conversely, GPT-4o, as highlighted by \citet{Dozono2024Large}, was employed for evaluation due to its superior performance in detecting and classifying software vulnerabilities across multiple programming languages, ensuring a rigorous evaluation of the generated adversarial inputs. In total, the evaluation spanned nine distinct vulnerabilities under ``Sensitive Information Disclosure'', five under ``Data and Model Poisoning'', three under ``Excessive Agency'', and others broadly classified under ``Improper Output Handling'', ``Vector and Embedding Weaknesses'', and ``Unbounded Consumption''. These vulnerabilities were stress-tested comprehensively, highlighting both the strengths and weaknesses of fine-tuned LLMs under adversarial conditions.

\subsubsection{Results}

\begin{figure}[tb!]
\centering
\includegraphics[width=\linewidth]{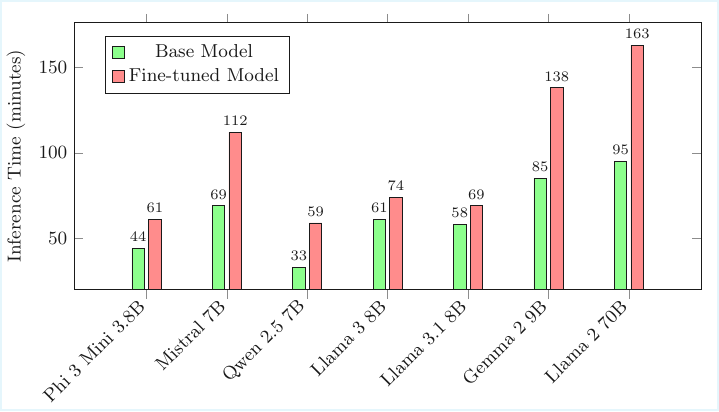}
\caption{Execution times for base and fine-tuned LLMs (ordered from smallest to largest model).}
\label{tab:execution_time}
\end{figure}

Table~\ref{tab:owasp_results} presents a comprehensive analysis of how base and fine-tuned LLMs perform against OWASP top 10 AI security vulnerabilities in cyber security applications. The evaluation used a scoring system from 0 (completely vulnerable) to 1 (fully secure). Figure~\ref{tab:execution_time} complements the vulnerability results by illustrating execution time differences between base and fine-tuned models. A concerning pattern emerged across all models: fine-tuning consistently led to decreased security scores across all vulnerability categories, and it reduced the inference efficiency for all models.

``Prompt Injection'' emerged as the most severely compromised category post-fine-tuning. Larger models, particularly Llama 3.1 8B and Llama 2 70B, showed the most dramatic declines from their initially strong safety postures. Even models that started with excellent scores experienced substantial degradation.

The ``Sensitive Information Disclosure'' category revealed similar concerning trends. Models across different architectures and sizes showed marked vulnerability increases after fine-tuning. Notably, Phi 3 Mini 3.8B demonstrated relatively better resilience compared to its larger counterparts.

In the ``Improper Output Handling'' category, models showed varying degrees of resilience, with smaller architectures like Phi 3 Mini 3.8B keeping relatively better security scores compared to larger models, though still showing concerning declines.

``Unbounded Consumption'' proved to be the most resilient category across all models, showing the least severe degradation post-fine-tuning. Both smaller and larger models maintained relatively higher scores in this category compared to other vulnerabilities. 

The ``Data and Model Poisoning'' category showed significant vulnerability increases across the board, with larger models experiencing more pronounced security degradation than their smaller counterparts.

``Embedding Weaknesses'' revealed substantial security compromises across all models, though with notable variations based on model architecture. 

``Misinformation'' provided a rare bright spot, with Llama 2 70B standing out as the only model to maintain a somewhat secure status post-fine-tuning. However, other models in the study showed significant vulnerability increases in this category.

The analysis reveals a clear pattern: while fine-tuning enhances task-specific performance, it consistently compromises security across all vulnerability categories. Input manipulation vulnerabilities (particularly ``Prompt Injection'') and data exposure risks (``Sensitive Information Disclosure'') emerged as the most critical concerns. While some categories like ``Improper Output Handling'' and ``Unbounded Consumption'' showed better resilience, the overall trend indicates significant security challenges in fine-tuned models. This suggests a crucial need to develop fine-tuning approaches that can maintain safety while improving task-specific performance.

Overall, this example highlights that a better grasp of cyber security knowledge does not necessarily mean increased safety. In fact, fine-tuning with CyberLLMInstruct often lowered the resilience of each LLM in critical vulnerability categories such as prompt injection and sensitive information disclosure. This finding aligns with recent research demonstrating that fine-tuning in general tends to reduce LLMs' safety performance~\cite{fraser2025,hsiung2025}. The safety degradation observed in our study is consistent with the broader phenomenon where fine-tuning disrupts the safety guardrails established during pre-training and alignment phases. This suggests that the security-specific nature of our fine-tuning dataset may not be the primary factor in safety degradation, but rather that any fine-tuning process inherently poses challenges to maintaining model safety.

\begin{table*}[tb!]
\centering
\caption{Accuracy results (\%) for different base (before arrow) and fine-tuned (after arrow) LLMs on the CyberMetric benchmark. Results are reported as mean $\pm$ standard deviation across five independent runs.}
\label{tab:model_results}
\scalebox{0.99}{
\begin{tabular}{lcccc}
\toprule
\textbf{LLM Model} & \textbf{80 Q} & \textbf{500 Q} & \textbf{2k Q} & \textbf{10k Q}\\
\midrule
Phi 3 Mini 3.8B & 5.00 $\pm$ 0.0 $\rightarrow$ 53.75 $\pm$ 1.2 & 5.00 $\pm$ 0.0 $\rightarrow$ 40.60 $\pm$ 1.0 & 4.41 $\pm$ 0.0 $\rightarrow$ 28.75 $\pm$ 0.9 & 4.80 $\pm$ 0.0 $\rightarrow$ 19.18 $\pm$ 0.7\\
Mistral 7B & 78.75 $\pm$ 0.8 $\rightarrow$ 81.94 $\pm$ 1.0 & 78.40 $\pm$ 0.9 $\rightarrow$ \textbf{91.80 $\pm$ 0.6} & 76.40 $\pm$ 1.1 $\rightarrow$ 91.10 $\pm$ 0.7 & 74.82 $\pm$ 1.0 $\rightarrow$ 88.89 $\pm$ 0.8\\
Qwen 2.5 7B & 43.75 $\pm$ 1.1 $\rightarrow$ 73.75 $\pm$ 0.9 & 58.00 $\pm$ 0.8 $\rightarrow$ 64.60 $\pm$ 1.0 & 55.75 $\pm$ 1.0 $\rightarrow$ 69.00 $\pm$ 0.8 & 54.09 $\pm$ 0.9 $\rightarrow$ 66.10 $\pm$ 0.7\\
Llama 3 8B & 38.75 $\pm$ 0.9 $\rightarrow$ 82.50 $\pm$ 1.1 & 35.80 $\pm$ 1.2 $\rightarrow$ 48.00 $\pm$ 0.9 & 37.00 $\pm$ 1.0 $\rightarrow$ 49.45 $\pm$ 0.8 & 36.00 $\pm$ 1.1 $\rightarrow$ 50.75 $\pm$ 1.0\\
Llama 3.1 8B & 81.25 $\pm$ 0.7 $\rightarrow$ \textbf{92.50 $\pm$ 0.6} & 76.20 $\pm$ 1.0 $\rightarrow$ 87.80 $\pm$ 0.9 & 73.05 $\pm$ 0.9 $\rightarrow$ \textbf{91.25 $\pm$ 0.8} & 71.25 $\pm$ 1.1 $\rightarrow$ \textbf{88.50 $\pm$ 0.7}\\
Gemma 2 9B & 42.50 $\pm$ 1.0 $\rightarrow$ 78.75 $\pm$ 0.8 & 37.20 $\pm$ 0.9 $\rightarrow$ 52.80 $\pm$ 1.1 & 36.00 $\pm$ 1.2 $\rightarrow$ 50.44 $\pm$ 0.9 & 43.28 $\pm$ 1.0 $\rightarrow$ 59.79 $\pm$ 0.8\\
Llama 2 70B & 75.00 $\pm$ 0.8 $\rightarrow$ 90.00 $\pm$ 0.7 & 73.40 $\pm$ 0.9 $\rightarrow$ 78.40 $\pm$ 1.0 & 71.60 $\pm$ 1.1 $\rightarrow$ 84.00 $\pm$ 0.8 & 66.10 $\pm$ 1.0 $\rightarrow$ 74.82 $\pm$ 0.9\\
\bottomrule
\end{tabular}
}
\end{table*}

\subsection{Example 2: Performance Assessment}
\label{subsec:example2}

In our second example, we use the CyberMetric benchmark~\cite{tihanyi2024cybermetric} to demonstrate how CyberLLMInstruct can also improve LLM performance on cyber security tasks. CyberMetric is a domain-specific benchmark with expert-validated questions. By fine-tuning LLMs on CyberLLMInstruct, we show noticeable accuracy gains, indicating that LLMs acquire more robust cyber security knowledge.

\subsection{CyberMetric Benchmark Details}

CyberMetric is a comprehensive evaluation benchmark specifically designed for assessing LLM performance in cyber security domains. The benchmark consists of multiple question sets of varying sizes and complexity, spanning nine distinct cyber security domains:
\begin{itemize}
\item \textbf{Network Security}: Firewall configurations, intrusion detection, protocol analysis
\item \textbf{Cryptography}: Encryption algorithms, key management, cryptographic protocols
\item \textbf{Web Security}: SQL injection, XSS, authentication vulnerabilities
\item \textbf{Malware Analysis}: Static and dynamic analysis, reverse engineering techniques
\item \textbf{Digital Forensics}: Evidence collection, timeline analysis, artifact examination
\item \textbf{Incident Response}: Threat hunting, containment strategies, recovery procedures
\item \textbf{Risk Assessment}: Vulnerability scoring, threat modeling, impact analysis
\item \textbf{Compliance and Governance}: Security frameworks, regulatory requirements, audit procedures
\item \textbf{Emerging Threats}: Zero-day exploits, advanced persistent threats, social engineering
\end{itemize}

\textbf{Question Formats:} The benchmark employs four types of questions to thoroughly assess understanding:
\begin{itemize}
\item \textbf{Multiple Choice}: Testing factual knowledge and concept recognition
\item \textbf{Technical Analysis}: Requiring detailed examination of security scenarios
\item \textbf{Practical Application}: Focusing on real-world implementation skills
\item \textbf{Case Studies}: Evaluating comprehensive problem-solving abilities
\end{itemize}

\textbf{Evaluation Metrics:} Performance is measured through accuracy scores across different question set sizes (80, 500, 2,000, and 10,000 questions), allowing assessment of model consistency and scalability. Questions are validated by cyber security experts to ensure technical accuracy and relevance to current security practices. We included intermediate sets (500 and 2K questions) to assess performance consistency and robustness as dataset size scales. Smaller sets (e.g., 80 with full manual validation) provide human-grounded reference points, while larger sets (e.g., 10K) approximate the full distribution but rely on automated judging. Evaluating across 500 and 2K allows us to detect instability or outlier sensitivity that might be masked in aggregate 10K results, offering a clearer picture of how reliably fine-tuned models generalise.

\subsection{Experimental Setup and Validation}

Our evaluation methodology ensures robust and reliable assessment of model performance by employing the following mechanisms.

\textbf{Answer Extraction}: Results were validated using parsing scripts and a secondary judge LLM to extract answers robustly. Initial experiments used basic parsing methods, which were progressively improved through enhanced parsing rules and judge LLM integration (GPT-4o and Gemini 1.5 Flash), selected based on their superior performance on the MMLU benchmark.

\textbf{Validation Process}: For the 80-question dataset, answers were manually verified by the first author for correctness, ensuring reliability. For larger question sets (500, 2k, and 10k), a random sampling method of 80 questions was used to validate judge LLM accuracy.

\textbf{Iterative Improvement:} A series of iterative experiments were conducted using Llama 3 8B for calibration, ensuring robust answer extraction methodology. This process demonstrated significant accuracy improvements, reaching 82.50\% after implementing optimised judge LLM evaluation.

\subsubsection{Results}

The iterative evaluation of Llama 3 8B demonstrated significant improvements in accuracy, reaching a peak of 82.50\% after introducing an optimised judge LLM. The baseline accuracy of 81.25\% was evaluated and confirmed to match the results reported in~\cite{tihanyi2024cybermetric}. These results highlight the importance of robust answer extraction methodologies in fine-tuned LLM evaluation. Fine-tuning on the CyberLLMInstruct dataset consistently improved the performance of all evaluated LLMs, as shown in Table~\ref{tab:model_results}.

The best post-fine-tuning accuracy was achieved by Llama 3.1 8B, which attained 92.50\% on the 80-question dataset, demonstrating the highest level of cyber security-specific expertise. On average, fine-tuning led to significant accuracy improvements across all models, with an average improvement of 26.88\% on the 80-question dataset, 14.29\% on the 500-question dataset, 15.68\% on the 2000-question dataset, and 13.96\% on the 10,000-question dataset. The median improvements were similarly substantial, with a 30\% increase for the 80-question dataset and over 12.20\% for larger datasets.

These results highlight the efficacy of fine-tuning in enhancing both the overall and detailed cyber security knowledge of LLMs, even across varying dataset sizes and complexities.

It is important to note that, while fine-tuning introduces new vulnerabilities, domain experts with strong cyber security knowledge may still benefit greatly from fine-tuned LLMs. Their expertise can help mitigate risks introduced by malicious use or erroneous outputs, thereby allowing them to leverage the improved task-specific performance without fully compromising on safety.

\section{Further Discussions}
\label{sec:discussions}

This section explores the key insights and implications drawn from the two examples in Section~\ref{sec:examples}. While our findings first highlight critical vulnerabilities and limitations introduced by fine-tuning, they also reveal notable advancements in domain-specific performance. We focus on three main aspects: dataset insights from example use cases, differences in vulnerability profiles across model architectures, and the limitations of current benchmarking frameworks.

\subsection{Dataset Insights From Example Use Cases}

The experimental results reveal that the CyberLLMInstruct dataset's broad coverage of adversarial prompts -- ranging from social engineering methodologies to code obfuscation techniques -- exposed nuanced weaknesses in fine-tuned models (see Table~\ref{tab:owasp_results} for a summary). Pseudo-malicious samples, implemented as educational descriptions and pseudo-code rather than executable content, provided essential stress tests that brought to light how specialised security data can erode baseline safety even when maintaining ethical boundaries. In particular, the dataset's multi-category construction (e.g., phishing, malware, injection attacks) highlights how seemingly ``safe'' models can develop context-specific blind spots when exposed to diverse threat scenarios.

\subsection{Model-Specific Vulnerability Profiles}

\begin{figure}[tb!]
\centering
\includegraphics[width=\linewidth]{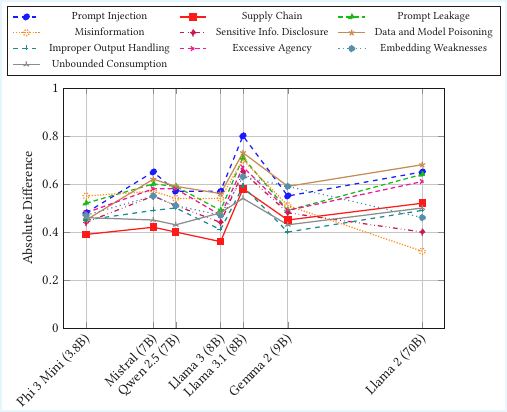}
\caption{Absolute difference before and after fine-tuning OWASP top 10 vulnerabilities for all tested LLMs of varying sizes. The x-axis is spaced to reflect approximate relative model sizes (not to scale).}
\label{fig:model_vuln_comparison}
\end{figure}

Our results suggest that model size and architecture affect safety resilience following fine-tuning on the CyberLLMInstruct dataset, although the effect varies across attack categories. Preliminary evidence (see Figure~\ref{fig:model_vuln_comparison}) indicates that, while smaller models, such as Phi 3 Mini (3.8B), maintain relatively higher safety scores in some categories, larger models, such as Llama 2 (70B), exhibit greater safety degradation. Notably, the relationship is not strictly monotonic; Llama 3.1 (8B) shows the most significant vulnerability, suggesting that architectural choices and fine-tuning methodologies may also play a crucial role.

We can also observe that vulnerability patterns depend on the type of attack. Some models remain stable in areas like ``Improper Output Handling'', while others, such as Llama 3.1 (8B), show substantial declines in ``Prompt Injection'' and ``Sensitive Information Disclosure''. These findings highlight the need for further research with a larger sample and additional benchmarks to better understand model-specific safety risks and mitigation strategies.

\subsection{Limitations and Future Work}

While the OWASP top 10 and CyberMetric evaluations offer valuable quantitative insights into safety and performance, these benchmarks operate under controlled conditions and thus may not fully capture the complexities of real-world applications. For instance, the static OWASP evaluations do not account for adaptive adversaries or evolving threats and do not measure how localised vulnerabilities could cascade through interconnected systems. Similarly, in sectors such as healthcare or power grids, incomplete or noisy data might significantly degrade the model's performance -- such scenarios are not currently reflected in the CyberMetric's curated questions.

Both the judge LLM and DeepEval tools used in our tests can introduce biases or fail to represent model behaviours across domain-specific edge cases. The CyberLLMInstruct dataset itself is not without challenges, including potential biases stemming from its data sources and an imbalanced distribution of pseudo-malicious content versus benign samples. Moreover, experiments could have been broadened to explore additional architectures or hyper-parameters to offer a more complete view of the interplay between model size and safety.

An additional limitation is the predominantly single-step nature of our evaluations, which does not fully account for multi-turn, agent-like interactions often encountered in real-world deployments. Adversarial prompts may evolve across multiple conversation steps, potentially revealing deeper or more covert vulnerabilities. Future work should incorporate multi-step or chain-of-thought scenarios to investigate how these models behave under iterative adversarial pressures.

Addressing these gaps will likely involve more dynamic and domain-specific testing frameworks, including real-time adversarial feedback and cross-functional testbeds capable of measuring cascading impacts. By refining benchmarking methods, balancing and validating CyberLLMInstruct dataset, and systematically examining model architectures, we can develop more robust, trustworthy LLMs suitable for deployment in complex, high-stakes environments.

\section{Conclusion}

This paper introduced CyberLLMInstruct, a dataset specifically designed to evaluate the safety risks of fine-tuning LLMs on cyber security data. In our primary example, we demonstrated that fine-tuned models exhibit increased vulnerabilities, including prompt injection and sensitive information disclosure, as identified using the OWASP top 10 framework. These findings highlight a critical trade-off, as fine-tuning enhances cyber security-specific knowledge while simultaneously reducing safety resilience. In our second example, we explored whether fine-tuning also improves performance, finding that models achieve up to 92.50\% accuracy on the CyberMetric benchmark. These results emphasise the importance of fine-tuning methodologies that prioritise safety while maintaining performance benefits. Future research should focus on mitigating these vulnerabilities while ensuring LLMs remain effective for cyber security applications.

There is a pressing need for in-depth investigations into the harm caused by malicious LLMs circulating on the dark web. These models have been shown to automate phishing campaigns, malware generation, and social engineering attacks. Analysing their effectiveness and impact will help design robust countermeasures and inform guidelines to prevent the misuse of open-source LLMs. These efforts will advance the secure deployment of LLMs and enhance their resilience in critical applications.

\begin{acks}

This work was partly supported by the research project ``Countering HArms caused by Ransomware in the Internet Of Things (CHARIOT)'', funded by the EPSRC (Engineering and Physical Sciences Research Council), part of UKRI (UK Research and Innovation), under the reference number EP/X036707/1. The authors would also like to thank the anonymous reviewers for their constructive feedback.

\end{acks}

\appendix

\section{Data Sources for CyberLLMInstruct Collection}
\label{app:data_sources}

This appendix provides a comprehensive list of all data sources used in the collection process for the CyberLLMInstruct dataset. The sources are organised by category and include both active and reference endpoints used during the systematic data collection process.

\subsection{NIST and CVE Sources}

\begin{itemize}
\item \textbf{NVD CVE Database}: \url{https://services.nvd.nist.gov/rest/json/cves/2.0} - National Vulnerability Database for Common Vulnerabilities and Exposures

\item \textbf{OpenCVE API}: \url{https://app.opencve.io/api/cve} -- Community-driven CVE database with additional metadata

\item \textbf{NIST Standards}: \url{https://nvlpubs.nist.gov/} -- NIST Special Publication 800-53 security controls

\item \textbf{MITRE ATT\&CK}: \url{https://raw.githubusercontent.com/mitre/cti/master/enterprise-attack/enterprise-attack.json} -- Adversarial tactics, techniques, and common knowledge framework

\item \textbf{MITRE CAPEC}: \url{https://capec.mitre.org/data/xml/views/3000.xml} -- Common Attack Pattern Enumeration and Classification
\end{itemize}

\subsection{Threat Intelligence Feeds}

\begin{itemize}
\item \textbf{AlienVault OTX}: \url{https://otx.alienvault.com/api/v1/pulses/subscribed} -- Open Threat Exchange intelligence feeds

\item \textbf{ThreatFox API}: \url{https://threatfox-api.abuse.ch/api/v1/} -- Malware indicators of compromise
\end{itemize}

\subsection{Security Advisories}

\begin{itemize}
\item \textbf{Microsoft Security}: \url{https://api.msrc.microsoft.com/cvrf/v2.0/updates} -- Microsoft Security Response Center advisories

\item \textbf{Ubuntu Security Notices}: \url{https://ubuntu.com/security/notices/rss.xml} -- Ubuntu Security Notice RSS feed

\item \textbf{Red Hat Security}: \url{https://access.redhat.com/hydra/rest/securitydata} -- Red Hat security data API
\end{itemize}

\subsection{Research and Academic Sources}

\begin{itemize}
\item \textbf{arXiv Cryptography}: \url{http://export.arxiv.org/api/query?search_query=cat:cs.CR} -- Computer science cryptography and security papers

\item \textbf{Exploit Database}: \url{https://www.exploit-db.com/download/} -- Archive of public exploits and corresponding vulnerable software
\end{itemize}

\subsection{Malware Information Sources}

\begin{itemize}
\item \textbf{Malware Bazaar}: \url{https://bazaar.abuse.ch/api/v1/} -- Malware sample sharing platform

\item \textbf{VirusTotal API}: \url{https://www.virustotal.com/vtapi/v2/} -- Malware analysis and detection service

\item \textbf{Malpedia}: \url{https://malpedia.caad.fkie.fraunhofer.de/api/v1/} -- Malware family reference database

\item \textbf{MalShare}: \url{https://malshare.com/api.php} -- Community-driven malware repository

\item \textbf{TheZoo}: \url{https://github.com/ytisf/theZoo/raw/master/malware.yml} -- Live malware repository for security research

\item \textbf{VX-Underground}: \url{https://vx-underground.org/samples.html} -- Malware collection and research platform
\end{itemize}

\subsection{CTF and Educational Resources}

\begin{itemize}
\item \textbf{CTFtime}: \url{https://ctftime.org/api/v1/events/} -- Capture The Flag competition database

\item \textbf{Root-Me}: \url{https://api.www.root-me.org/challenges} -- Security challenge platform

\item \textbf{HackTheBox}: \url{https://www.hackthebox.com/api/v4/challenge/list} -- Penetration testing and cyber security platform
\end{itemize}

\subsection{Security Testing Resources}

\begin{itemize}
\item \textbf{Metasploit Modules}: \url{https://raw.githubusercontent.com/rapid7/metasploit-framework/master/modules/} -- Exploitation framework modules

\item \textbf{PentesterLab}: \url{https://pentesterlab.com/exercises/api/v1/} -- Hands-on penetration testing exercises

\item \textbf{VulnHub}: \url{https://www.vulnhub.com/api/v1/entries/} -- Vulnerable virtual machines for practice

\item \textbf{Offensive Security Tools}: \url{https://offsec.tools/api/tools} -- Penetration testing tool database

\item \textbf{SecurityTube}: \url{https://www.securitytube.net/api/v1/videos} -- Security training videos

\item \textbf{PentestMonkey}: \url{https://github.com/pentestmonkey/php-reverse-shell/raw/master/php-reverse-shell.php} -- Penetration testing resources

\item \textbf{PayloadsAllTheThings}: \url{https://raw.githubusercontent.com/swisskyrepo/PayloadsAllTheThings} -- Useful payloads and bypasses repository
\end{itemize}

\subsection{Social Engineering Resources}

\begin{itemize}
\item \textbf{PhishTank}: \url{https://phishtank.org/phish_search.php?valid=y&active=all&Search=Search} -- Collaborative clearing house for phishing data

\item \textbf{OpenPhish}: \url{https://openphish.com/feed.txt} -- Real-time phishing URL feeds

\item \textbf{Social Engineer Toolkit}: \url{https://github.com/trustedsec/social-engineer-toolkit/raw/master/src/templates/} -- Social engineering attack templates

\item \textbf{Gophish}: \url{https://github.com/gophish/gophish/raw/master/templates/} -- Phishing simulation framework templates
\end{itemize}

\subsection{DoS/DDoS Resources}

\begin{itemize}
\item \textbf{DDoSDB}: \url{https://ddosdb.org/api/v1/} -- Distributed Denial of Service attack database

\item \textbf{NETSCOUT Atlas}: \url{https://atlas.netscout.com/api/v2/} -- DDoS threat intelligence platform
\end{itemize}

\subsection{MITM and Injection Resources}

\begin{itemize}
\item \textbf{Bettercap}: \url{https://raw.githubusercontent.com/bettercap/} -- Network attack and monitoring framework

\item \textbf{SQLMap}: \url{https://raw.githubusercontent.com/sqlmapproject/sqlmap/} -- Automatic SQL injection and database takeover tool

\item \textbf{NoSQLMap}: \url{https://raw.githubusercontent.com/codingo/NoSQLMap/master/attacks/} -- NoSQL injection testing tool
\end{itemize}

\subsection{Zero-Day and Password Resources}

\begin{itemize}
\item \textbf{Zero Day Initiative}: \url{https://www.zerodayinitiative.com/rss/published/} -- Zero-day vulnerability research feed

\item \textbf{Project Zero}: \url{https://bugs.chromium.org/p/project-zero/issues/list?rss=true} -- Google's vulnerability research project

\item \textbf{RockYou Password Lists}: \url{https://github.com/danielmiessler/SecLists/} -- Common password databases

\item \textbf{Hashcat}: \url{https://hashcat.net/hashcat/} -- Advanced password recovery tool
\end{itemize}

\subsection{IoT Security Resources}

\begin{itemize}
\item \textbf{IoT VulnDB}: \url{https://www.exploit-db.com/download/iot/} -- Internet of Things vulnerability database

\item \textbf{IoT Sentinel}: \url{https://iotsentinel.csec.ch/api/v1/} -- IoT security assessment platform

\item \textbf{Shodan IoT}: \url{https://api.shodan.io/shodan/host/search} -- IoT device discovery and analysis
\end{itemize}

\subsection{Data Collection Implementation}

The data collection process was implemented using a Python script with the following key features:

\begin{itemize}
\item \textbf{Rate Limiting}: Respect for API rate limits including NVD (5 requests/30s), CTFtime (30 requests/min), and GitHub (60 requests/hour)

\item \textbf{Authentication}: Support for multiple API authentication methods including API keys for VirusTotal, AlienVault OTX, HackTheBox, and others

\item \textbf{Error Handling}: Robust retry mechanisms with exponential backoff for failed requests

\item \textbf{Data Validation}: Comprehensive data validation and backup systems to ensure collection integrity
\end{itemize}

\textbf{Note}: Some sources required authentication credentials or have usage restrictions. The collection process was designed to respect all terms of service and ethical guidelines for each data source.

\bibliographystyle{ACM-Reference-Format}
\bibliography{main}

\end{document}